\documentclass[structabstract]{aa}
\usepackage{graphicx}
\usepackage{txfonts}
\usepackage{natbib}
\bibpunct{(}{)}{;}{a}{}{,}

\begin{document}

\def\clgs{ClG 0332-2747}
\def\S09{S09}
\title{An X-ray underluminous cluster of galaxies in the 4Ms CDFS observations}

   \author{M. Castellano \inst{1}
 \and
     L. Pentericci
     \inst{1}
  \and
    N. Menci
     \inst{1}
  \and
    E. Piconcelli
     \inst{1}
  \and
    P. Santini
     \inst{1}
  \and
    S. Salimbeni
     \inst{2}
 \and
     F. Fiore
      \inst{1}
   \and
     A. Fontana
      \inst{1}
     \and
     A. Grazian
     \inst{1}
 \and
     A. Romano
     \inst{1}
  \and
      D. Trevese
\inst{3}
   }

   \offprints{M. Castellano, \email{marco.castellano@oa-roma.inaf.it}}

\institute{INAF - Osservatorio Astronomico di Roma, Via Frascati 33,
I--00040 Monteporzio (RM), Italy
  \and Department of Astronomy, University of Massachusetts, 710 North Pleasant Street, Amherst, MA 01003, USA \and Dipartimento di Fisica,
 Universit\`{a} di Roma ``La Sapienza'', P.le A. Moro 2, 00185 Roma, Italy} 

   \date{Received .... ; accepted ....}
   \titlerunning{An X-ray underluminous cluster in the 4Ms CDFS}

\abstract
{}
{
We investigate the properties of cluster \clgs~at z=0.734 in the GOODS-South field, which was undetected in the Chandra 2Ms observations. We explore possible scenarios to explain  the discrepancy between its low X-ray emission and that 
 expected from the $M-L_X$ relation. 
}
{We use the large  public spectroscopic database available in the GOODS-South field to estimate the dynamical mass and the virialization status of \clgs. Cluster members selected from their photometric redshift are used with spectroscopic ones to analyse the galaxy population of the cluster. In the newly released Chandra 4Ms observations we detect a faint extended X-ray emission associated to the cluster. Finally, we compare the optical and X-ray properties of \clgs~with the predictions of a well-tested semianalytic model. 
}
{We estimate the velocity dispersion and the virial mass considering all 44 spectroscopic members, or 20 red-sequence members only. 
We obtain  $\sigma_v=634 \pm 105~ Km/s$,
$M_{200}=3.07 ^{+1.57} _{-1.16}~10^{14}~ M_{\odot}$  in the former case, and slightly lower values in the latter case. The cluster appears to have reached the virial equilibrium: it shows a perfectly Gaussian velocity distribution and no evidence for substructures. \clgs~contains a high fraction of bright red galaxies and is dominated by a very massive ($1.1\times10^{12}~ M_{\odot}$) old brightest cluster galaxy (BCG), suggesting that it formed at an early epoch. We detect a faint extended X-ray source centred on the BCG, with a total X-ray luminosity of $L_X \sim 2\cdot10^{42}~erg ~s^{-1}$ (0.1-2.4 keV). This $L_X$ is lower by a factor of $\sim$10-20 than expected according to the $M-L_X$ relation. We provide a possible explanation of this discrepancy assigning it to the effects of AGN feedback on the ICM: the semianalytic model reproduces the $M-L_X$ relation measured from ``X-ray bright'' clusters, and it predicts a high scatter at low masses owing to heating and expulsion of the cluster gas. Interestingly, the model clusters with an evolved galaxy population like \clgs~present the largest scatter in X-ray luminosity. However, the low X-ray emission of \clgs~is just marginally compatible with predictions, which indicates that additional feedback effects should be included in the model. }
{We propose a scenario where ``X-ray underluminous'' clusters are explained by the strong feedback effect on the ICM in highly evolved clusters. Our hypothesis can be tested by the combined analysis of the galaxy population and of the X-ray emission in large cluster samples.}


\maketitle
%
\section{Introduction}

A multiwavelength approach in the detection and investigation of distant galaxy clusters is necessary to determine the interconnection between the evolution of the dark matter component and of their baryon content, and to determine the impact of selection biases in available and future cluster samples. The area that can probably benefit most from this approach is the study of the shape and evolution of cluster scaling relations. Assuming that gravity alone determines the evolution of their gas halos, galaxy clusters of different masses are expected to be scaled versions of each other (i.e. to be ``self-similar''), and to show simple power-law relations between their properties $M_{tot}$, $M_{gas}$, $L_X$, $T$ \citep{Kaiser1986}. It was soon realised that the observed scaling relations can have different forms with respect to simple self-similar predictions \cite[e.g.][]{Markevitch1998,Reiprich2002}; in addition it is still under debate whether evolution with redshift has a self-similar beahviour \cite[e.g.][]{Ettori2004,Maughan2007,Pacaud2007}.

The observed deviations from self-similarity arise because of non-gravitational processes, either through a pre-heating mechanism \citep{Kaiser1991} or through the internal impact of feedback from supernovae \citep{Borgani2005,Muanwong2006} and active galactic nuclei (AGN) \citep{Puchwein2008,McCarthy2010,Short2010}. In this respect, optical analyses are needed to properly assess the physical origin of these deviations since optical mass estimates directly trace the depth of the cluster potential well, either through galaxy dynamics or weak lensing. When compared to X-ray estimates they allow us to directly evaluate the non-gravitational processes acting on the gas. 
The reported differences in the scaling relations between optically and X-ray selected clusters \cite[e.g.][]{Bignamini2008,Hicks2008} further indicate that unbiased samples are needed to investigate the extent to which non gravitational processes modify cluster gas properties.

Clusters that show striking discrepancies between X-ray and optical properties are optimal targets for investigating the impact of non-gravitational effects in the evolution of the intracluster medium. This is the case of the so-called ``X-ray underluminous'' clusters, which exhibit much lower X-ray luminosities than expected from their optically estimated mass, velocity dispersion, or richness \cite[e.g.][]{Bower1994,Bower1997,Lubin2002,Lubin2004,Gilbank2004,Hicks2004,Rasmussen2006,Fang2007,Popesso2007}.

Indeed, their low X-ray luminosity might be owing to a small
gas content, either because feedback ejected a significant fraction of the ICM \citep{Rasmussen2006}, or because of a higher efficiency of galaxy formation \citep{Gilbank2004}. Alternatively it has been suggested that they might still be in a forming process \citep{Lubin2002,Rasmussen2006,Popesso2007,Dietrich2009}.  On the other hand, optical estimates can be plagued by difficulties in separating virialized and infalling cluster members \citep{Bower1997}, and in classyfing mergers, chance projections or filaments observed along the line of sight \citep{Bower1997,Gilbank2004,Lubin2004}. In this respect, \citet{Andreon2008} revised some claimed X-ray underluminous clusters, finding that most of them are indeed compatible with known scaling relations.

In our previous paper \cite[][hereafter \S09]{Salimbeni2009}, we presented a catalogue of 12 groups and small clusters at $0.4<z<2.5$ in the GOODS-South field detected through our (2+1)D algorithm \cite[described in detail in ][]{trevese2006}. 
On the basis of the then available Chandra 2Ms observations we found that the two most massive clusters in our sample at z=0.73 (cluster \clgs~hereafter) and at z=1.61 (ClG 0332-2742) were both undetected, implying an X-ray luminosity significantly lower than what is expected from the $M-L_X$ relation. 
While cluster ClG 0332-2742 at z=1.61 probably has not yet reached a relaxed status \citep{castelllano2007,Kurk2009}, cluster \clgs~appears to be very similar to lower redshift relaxed clusters. 

Here we present a more detailed analysis of cluster \clgs~and a comparison of its observed properties with the predictions of our semianalytic model. In particular, the present analysis is based on the large amount of new spectroscopic data on the GOODS-South field published in the last two years, and on the newly released Chandra 4MS observations of the CDFS.

The paper is organised as follows: in Sect.~\ref{data} we describe the data set. In Sect.~\ref{optical} we discuss the optical properties of  \clgs: velocity dispersion, virialization status, virial mass, total stellar mass and the properties of its galaxy population.  The X-ray analysis of the cluster, along with a comparison with the latest estimate of the $M-L_X$ relation, is discussed in Sect.~\ref{xray}, while a discussion of the possible physical origin of its properties is given in Sect.~\ref{discussion}. Finally, we summarize our results in Sect.~\ref{summary}.

All magnitudes used in the present paper are in the AB system,
if not otherwise declared. An $\Omega_\Lambda=0.7$, $\Omega_M=0.3$,
and $H_0=70$ km s$^{-1}$ Mpc$^{-1}$ cosmology is adopted.

\section{Data}\label{data}

We used the multicolour GOODS-MUSIC catalogue \citep[GOODS MUlticolour Southern Infrared Catalog;][]{grazian,Santini2009,Santini2009Cat}, which comprises information in 15 bands (from U band to 24$\mu m$) over an area of about 143.2 $arcmin^2$ in the Chandra Deep Field South.
The spectroscopic sample included in GOODS-MUSIC comes from several surveys \citep{wolf2001,lefevre2004,szokoly2004,Daddi2005,mignoli2005,vanzella2005,vanzella2006,vanzella2007,Popesso2009}, with the addition of  the latest redshifts in the V2.0.1 VIMOS release of \citet{Balestra2010}. 
The remaining objects in the catalogue have accurate photometric redshifts with $\sigma_{\Delta z/(1+z)}=0.03$ up to redshift $z=2$.

The X-ray analysis is based on the newly released 3.8Ms Chandra observation of the CDFS\footnote{http://cxc.harvard.edu/cda/Contrib/CDFS.html} built from the original exposures \citep{Giacconi2002,Luo2008} plus the 31 observations taken in 2010.

\section{Optical analysis}\label{optical}
We initially identified cluster \clgs~in \citet{trevese2006} by applying our algorithm to the data from the K20 catalogue, and we discussed its optical properties: it presents a well defined red sequence and an evident segregation of red galaxies typical of relaxed clusters.  In \S09 we estimated a mass in the range $0.9-3.0 ~10^{14}~ M_{\odot}$ (for bias factor b=2 and b=1 respectively) from the overdensity of galaxies, following the method by \citet{steidel98}. We also used the spectroscopic members of the overdensity to estimate a $\sigma_v\sim634 Km/s$ and $M_{vir}\sim3.2\cdot 10^{14} M_{\odot}$. 
Here we exploit all the new spectroscopic information available to re-estimate cluster position and virial mass, and to analyse the virialization status of the cluster in a self-consistent way.

We follow the method described by \citet{Biviano2006} (B06 hereafter) with minor adaptations to our data-set to compute velocity dispersion and virial mass for \clgs. After constraining the cluster extension in redshift space and selecting its spectroscopic members, we apply the two mass estimators proposed by B06: the classical virial mass estimator $M_{200}$ ($M_v$  in their paper) and their $M_{\sigma}$ estimator. The first one is based both on the velocity dispersion and on the harmonic mean radius of the cluster galaxies ($M_{200}\propto \sigma^2 R$); $M_{\sigma}$ comes from the best-fitting relation between velocity dispersion and mass in their simulated clusters. We explore possible systematics by considering all galaxies in the cluster and red-sequence members only.  The virialization status of the cluster is discussed in Sect. \ref{virstatus}. A summary of the properties of the cluster is given in Table ~\ref{tab:clust}. 
\begin{figure}
\includegraphics[height=8cm]{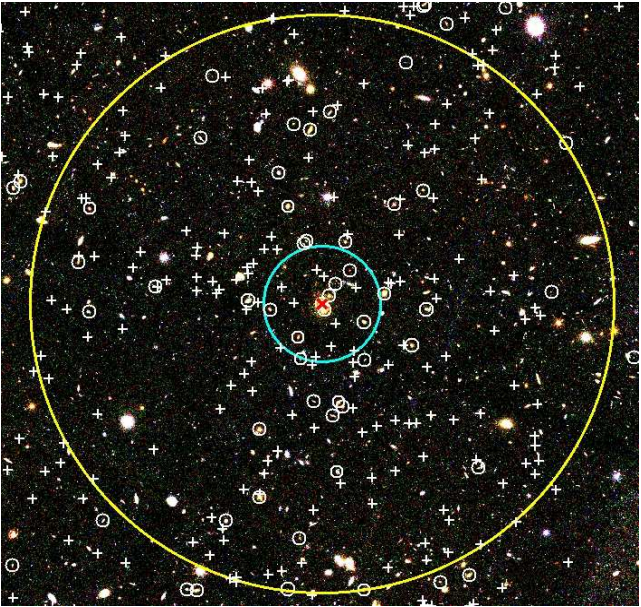}
\caption{RGB image of cluster \clgs~ at z=0.7347 in the GOODS-South field obtained from the V, I, Z HST images. The red x symbol indicates the cluster BCG, the large yellow circle encloses the area within 1 Mpc from the centre, the cyan circle indicates the area of radius 0.2 Mpc within which the X-ray emission is detected. Open circles mark the position of cluster spectroscopic members, crosses indicates all galaxies with photometric redshift within 2$\sigma_{photo-z}$ from the cluster redshift.}
\label{fig:fig_clgs}
\end{figure}
 \begin{figure}
 \includegraphics[height=8cm]{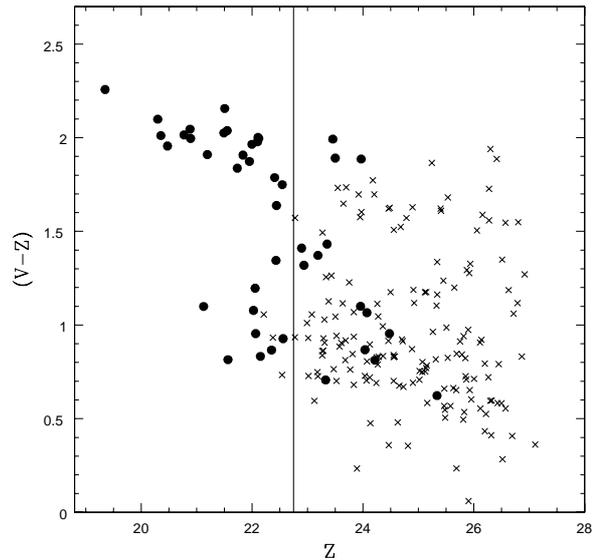}
 \caption{Observed colour magnitude diagram of \clgs~of all spectroscopic (black circles) and photometric (crosses) members of the cluster within 1 Mpc from the BCG. The vertical line at the observed Z=22.75 corresponds to V=-20.0 rest frame. The fraction of red galaxies (V-Z$\geq1.7$) brighter than this limit is $\sim$60\% within 1 Mpc and $\sim$74\% within $0.6\times R_{200}$.}
 \label{fig:colmag}
 \end{figure}

\subsection{Selection of cluster members}\label{clustercenter} We first constrained the cluster extension in redshift space through the ``gapper method'' described in \citet{beers1990}. We selected all spectroscopic galaxies within one Abell radius (2.15 Mpc) from the centre of the overdensity and in a wide redshift range ($\sim$20000 Km/s) around the known redshift peak of the structure  ($z\sim0.734$). By adopting a threshold of 4 in the normalized weighted gap distribution \cite[as suggested by][]{Girardi1993} we confined the cluster redshift extension to the interval $0.7286\leq z\leq0.7420$. This interval includes 97 spectroscopic members. We checked that this value does not change when adopting a more precise determination of the cluster centre as discussed below.

Among these members we searched for the brightest cluster galaxy (BCG) of \clgs.
Indeed, the best optical indicator of a cluster centre is the position of its central dominant galaxy, which is known to lie close to the peak of the emission in X-ray bright clusters \citep[e.g.][]{Jones1979}, and to be located at the bottom of the cluster potential well \citep[e.g.][]{Oegerle2001}. The BCG of cluster \clgs~ has been readily identified as the object $ID_{GOODS-MUSIC}=9792$ at RA=53.07506$^{\circ}$, DEC=-27.7885$^{\circ}$, which is the brightest galaxy in the sample, is a red-sequence member and lies at the peak of the spectroscopic redshift distribution.  
No interlopers in the projected phase-space were found by applying the method by \citet{Katgert1996} \cite[see also][]{Katgert2004} to all galaxies in the above redshift interval and up to one projected Abell radius from the BCG.

From the previous estimates of velocity dispersion for \clgs~ (\S09), and the $R_{vir}-\sigma_v$ relation in \citet{girardi1998}, we expect the virial radius, which separates the relaxed region from the infall region of the cluster \citep{Cole1996}, to be $\sim 1$ Mpc. In the following we  therefore estimate velocity dispersion and virial mass from the 44 spectroscopic members within 1.0 Mpc from the BCG.
To study the properties of the galaxy population in \clgs, we also exploit the 163 photometric members of the cluster including all galaxies within 1.0 Mpc from the BCG and having photometric redshift $0.63\leq z_{phot} \leq0.83$ ($\sim2\sigma_{photo-z}$ from the cluster redshift).

In Fig.~\ref{fig:colmag} we show the resulting colour-magnitude diagram of \clgs~in the observed ACS $V_{606}$ and $z_{850}$ bands (V and Z hereafter) bracketing the 4000\AA~break: the evident red sequence has a slope that is consistent with the one measured in low-redshift clusters, as discussed in \citet{trevese2006}.

\subsection{Velocity dispersion}
We used the biweight estimator \citep{beers1990} to compute the projected velocity dispersion $\sigma_{v}$, and we took into account redshift uncertainties following \citet{Danese1980}. The spectroscopic sample of \clgs~is based on observations carried out with different instruments, yielding redshift uncertainties ranging from $\delta z=$0.00040 \cite[VIMOS MR grism,][]{Balestra2010} up to $\delta z=$0.0012 \cite[VVDS,][]{lefevre2004}. We thus corrected the velocity dispersion according to the average redshift uncertainty of the sample, $\delta z=$0.001.

\begin{figure}
\includegraphics[height=8cm]{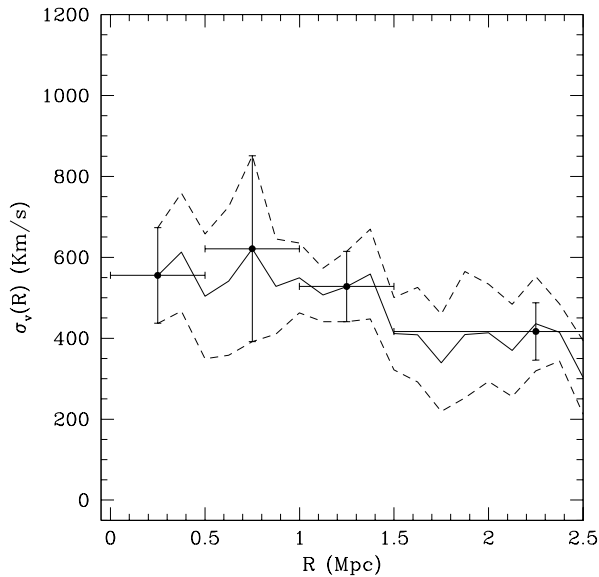}
\caption{Velocity dispersion profile of \clgs~ in four radial bins (black points and error bars) and computed through a moving window of 0.5 Mpc at steps of 0.125 Mpc (continuos and dashed lines).}
\label{fig:fig_sigmav_profile}
\end{figure}

The biweight estimate of the cluster mean redshift and the corrected velocity dispersion for all 44 spectroscopic members within 1 Mpc from the cluster center are $z=0.7347 \pm 0.0006$, $\sigma_{v}=634 \pm 105~Km/s$.  Uncertainties (68\% c.l.) were computed with the jackknife method as suggested by \citet{beers1990}. The velocity dispersion is perfectly consistent with the previous analysis of \S09. The velocity dispersion profile of \clgs, shown in Fig.~\ref{fig:fig_sigmav_profile}, appears nearly flat within 1 Mpc in agreement with an isotropic velocity distribution \citep{girardi1998}.

We also computed the velocity dispersion considering red spectroscopic members only. We applied the colour cut $V-Z\geq$1.7 to select all reliable red-sequence members, and we excluded the three brightest ones because they might have been slowed down by segregation effects \citep{Biviano1992}. From the 20 remaining objects we obtain a velocity dispersion $\sigma_{v,red}=516\pm118~Km/s$, which is consistent at $1\sigma$ with the velocity dispersion of the full sample, although slightly lower.

\subsubsection{Virialization status}
\label{virstatus}

\begin{figure}
\includegraphics[height=8cm]{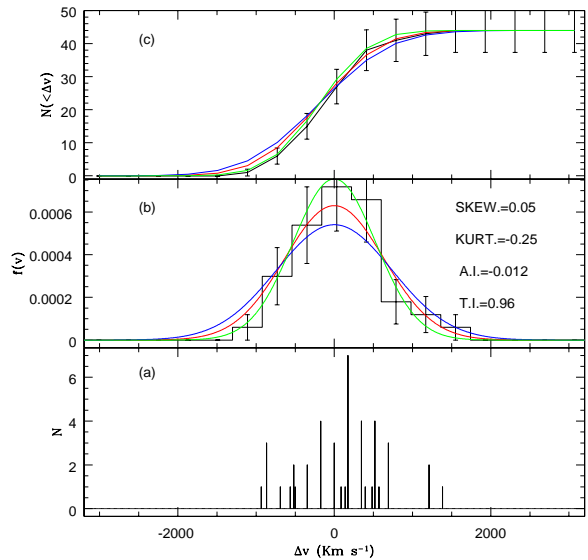}
\caption{\textit{Panel (a)}: rest-frame velocities of the 44 spectroscopic members of \clgs~ within 1.0 Mpc from the BCG.; \textit{Panel (b)}: binned velocity distribution, compared to Gaussians with dispersion obtained through the biweight estimate (red) and considering the jackknife uncertainties (blue and green). All distributions are normalized to 1.0; \textit{Panel (c)}: cumulative velocity distributions, colour code as in panel (b).}
\label{fig:fig_gauss}
\end{figure}

The velocity dispersion can be effectively used to estimate the total dynamical mass only if the cluster has reached a relaxed status (virial equilibrium).
This status, which is acquired through the process of ``violent relaxation'' \citep{LyndenBell1967},  is characterised by a Gaussian galaxy velocity distribution \cite[e.g.][]{Nakamura2000,Merrall2003} and, as shown by N-body simulations, by a low mass fraction included in substructures \cite[e.g.][]{Shaw2006}.
We performed five one-dimensional statistical tests to investigate whether the velocity distribution of the galaxy members is consistent with being Gaussian: the Kolmogorov-Smirnov test \cite[as implemented in the ROSTAT package of][]{beers1990}, two classical normality tests (skewness and kurtosis) and the two more robust asymmetry index (A.I.) and tail index (T.I.) described in \citet{Bird1993}. The expected values for a Gaussian distribution are 0.0 for skewness, kurtosis, and A.I. and 1.0 for the (normalized) T.I.. We find consistency on the basis of the K-S test, and values of 0.05, -0.25, -0.012, 0.96 for skewness, kurtosis, A.I., and T.I. respectively. By comparing these values with the confidence intervals reported in \citet{Bird1993} for a sample with 50 objects, we find that all the tests are perfectly consistent with the velocity distribution of \clgs~ being Gaussian. We plot in Fig.~\ref{fig:fig_gauss} the rest frame velocities of the 44 objects in our sample (panel a), the differential (b) and cumulative (c) velocity distributions, along with the best-fit Gaussian. The velocity distributions of ``red'' and ``blue'' galaxies, defined according to their (V-Z) colour as discussed above, are also separately consistent with being Gaussians.

We then performed the two-dimensional $\Delta$-test of \citet{Dressler1988} to look for substructures. In brief, this test quantifies the degree of correlation between the projected position and the velocity of each cluster member to locate cooler or hotter systems within the cluster, such as substructures or flows of infalling galaxies. We computed the individual $\delta$ values for each galaxy considering its $\sqrt{N_{members}}\simeq7$ nearest neighbours, as suggested by \citet{Pinkney1996}. The statistics was normalized by randomly shuffling galaxy velocities through 10000 Monte Carlo simulations. We obtain $P(\Delta)=0.254$, which is higher than the limit $P(\Delta)=0.1$ commonly adopted as indicative of substructures \cite[e.g.][]{Popesso2007}.
As a final test of the reliability of the estimated $\sigma_{v}$, we removed eight objects with $\delta>$3 as possible interlopers or infalling objects and computed the velocity dispersion of the remaining 36 cluster members (having $P(\Delta)\sim0.6$). We find a biweight $\sigma_{v}=600 \pm 113~ Km/s$, consistent with the one derived from the full sample.
\begin{table*}[]
\caption{Properties of \clgs} \label{tab:clust}
\begin{center}
\begin{tabular}{cccccccccc}
\noalign{\smallskip}\hline\noalign{\smallskip}\hline\noalign{\smallskip}
R.A.&DEC & Redshift  & $\sigma_v^a$& $\sigma_{v,red}^b$ & $M_{200}^a$ & $M_{200,red}^b$ & $M_{star}$ & Flux$^c$ & $L_{X}^d$ \\
 &  &  & & &  &  &  & (0.5-2 keV)  &(0.1-2.4 keV)  \\
 deg & deg & &Km/s  & Km/s & $10^{14}~ M_{\odot}$& $10^{14}~ M_{\odot}$ &  $10^{12}~ M_{\odot}$  & $10^{-16} \ erg \ s^{-1} \ cm^{-2}$ &$ 10^{42}\ erg \ s^{-1}$  \\
\noalign{\smallskip}\hline\noalign{\smallskip}
 53.07506 & -27.7885 & 0.7347 & $634 \pm 105$  & $516\pm118$ & $3.07 ^{+1.57} _{-1.16}$& $1.6 ^{+1.16} _{-0.8}$ & $3.3 \pm 0.5$& $2.3 \pm 0.7$&$2.1^{+0.69}_{-0.64}$\\
\noalign{\smallskip} \hline
\end{tabular}
\\
\medskip
\begin{tabular}{l}
a - From all 44 spectroscopic members within 1.0 Mpc from the BCG \\
b - From 20 red sequence members \\
c - Assuming $kT=3\pm1$ keV \\
d - Total X-ray luminosity assuming an average isothermal profile for clusters with $kT=3\pm1$ keV (Sect.~\ref{xray}). \\
\end{tabular}
\\
\end{center}
\end{table*}

\subsection{Virial mass}
Given the apparent relaxed status of \clgs~we used its velocity dispersion to estimate the virial mass $M_{200}$. We first estimated the cluster mass inside the aperture $a=$1.0 Mpc as $M_{a}\equiv M(r<a)=(3 \pi /2) \sigma_{v}^2 R_{PV} /G$, where $R_{PV}$ is the projected virial radius estimated from the distance between pairs of all galaxies within $r=a$ \citep[see ][for details]{girardi1998}. We corrected $M_{a}$ for the surface pressure term \citep{The1986} by applying a typical correction at $R_{vir}$ of 20\% as indicated by \citet{girardi1998} for isotropic galaxy orbits. We estimated the virial radius following \citet{carlberg} as $R_{200}=a\cdot [ \rho_a / (200 \rho_c (z)) ]^{1/\xi} $, where $\rho_a = \check{M_{a}}/(4 \pi a^3 /3)$, $\check{M_{a}}$ is the corrected value of $M_{a}$, $\rho_c (z)$ is the critical density of the Universe, and $\xi=2.4$ is the local slope of the cluster mass density profile near $R_{200}$ \citep{Katgert2004}. Finally, we extrapolated $\check{M_{a}}$ to $R_{200}$ considering a NFW profile \citep{Navarro1997} with concentration parameter $c=4.0 \cdot (\sigma_v/700.0)^{-0.306}$ \citep{Katgert2004}.

We obtain $R_{200}= 1.06 ^{+0.14} _{-0.15}~Mpc$, and $M_{200}= 3.07 ^{+1.57} _{-1.16} ~10^{14}~ M_{\odot}$ in agreement with the $M_{200}\sim 3.0 ~10^{14}~ M_{\odot}$ that we obtained in \S09 from the overdensity assuming that galaxies are unbiased tacers of the underlying dark matter distribution.

The virial mass computed through the $M_{\sigma}$ estimator of B06, which they find to be more robust against biases, is $M_{\sigma}=3.10 ^{+2.1} _{-1.5} ~10^{14}~ M_{\odot}$, consistent with the previous one. 

On the other hand, from the red galaxies sample we obtain a virial mass $M_{200,red}= 1.6^{+1.16} _{-0.8} ~\cdot 10^{14} ~M_{\odot}$ ($R_{200,red}= 0.86 ^{+0.16} _{-0.17}~Mpc$) and a consistent value for $M_{\sigma,red}$. 
The estimated $M_{200}$ and $M_{200,red}$ masses change by less than $\sim5\%$ varying the NFW concentration parameter in the range c=2-8, or the slope at $R_{200}$ in the range $\xi=2.0-2.8$. 

The difference between masses obtained with the full sample and those from red galaxies indicates that at least one of the two samples is a biased tracer of the cluster dynamics. We follow the discussion in B06 to interpret this discrepancy. They find that $M_{200}$ can be biased by the overestimate of the harmonic mean radius of the cluster due to the presence of interlopers in the sample. In their simulations, when estimating $M_{200}$ with red galaxies only, the presence of interlopers is reduced but not cancelled, which implies that the radius is still slightly overestimated. However, since early type galaxies underestimate the velocity dispersion, the resulting $M_{200}$ is closer to the true value.

On the other hand, $M_{\sigma}$ is less affected by the presence of interlopers being only based on the cluster velocity dispersion. This implies that $M_{\sigma}$ estimated from all cluster galaxies is more robust against biases, while it is biased low when using early type galaxies only. In our case, the agreement we find between the two estimators (in both samples) implies that $M_{200}$ and $M_{200,red}$ are unaffected by the presence of interlopers. This is also suggested by the results of the Dressler-Schechtman tests presented in the previous section. Thus, the lower value of $M_{200,red}$ is most probably caused by red galaxies underestimating the velocity dispersion of the cluster. In the following we will consider $M_{200}$ as the reference estimate, and we will keep $M_{200,red}$ as a lower limit for the cluster mass.

\subsection{Total stellar mass and fraction of red galaxies in \clgs}\label{stellarmass}
The properties of the galaxy population in clusters are linked to the formation history of the structure and show correlations to its total mass. 
We thus estimated the total stellar mass and investigated the properties of the BCG and of red sequence galaxies in \clgs~to provide a comparison with other high-redshift clusters.
The rest-frame magnitudes and stellar masses for galaxies in the GOODS-MUSIC sample were obtained through a SED fitting technique, as described in detail in \citet{grazian} and \citet{fontana06}. 

To compute the total stellar mass of the cluster we considered cluster members as defined in Sect~\ref{clustercenter}. 
All 28 massive objects ($M>10^{10}M_{sun}$) in this sample are bright spectroscopic members of the cluster, yielding a stellar mass $\sim 3 \cdot 10^{12}~ M_{\odot}$. We fitted a Schechter mass function (MF) considering both photometric and spectroscopic members through a maximum likelihood technique \cite[e.g.][]{Fontana2004}. Extrapolating this MF to lower masses, the resulting total stellar mass of the cluster is $M_{star}=3.3 \pm 0.5~ 10^{12}~ M_{\odot}$, which accounts for the uncertainties both on the SED fitting masses and on the MF parameters. 

A comparison with the cluster sample of \citet{Andreon2008} shows that the total stellar mass of \clgs~agrees with the one expected for clusters of $M_{200}\sim 1-3\cdot 10^{14}~ M_{\odot}$.  

The BCG, with a stellar mass $M=1.1 \cdot 10^{12} ~M_{\odot}$, contributes about one third to the total stellar mass. The BCG is probably also an AGN host because it is detected both in the radio \citep{Kellermann2008,Middelberg2011} and in X-ray \citep{Luo2008} while it shows no signs of ongoing star formation from its SED. It is associated to the point-like X-ray source 173 in the CDFS 2Ms catalogue by \citet{Luo2008}, with an X-ray 0.5-2 keV flux of $1.6\cdot10^{-16}~erg/cm^{2}/s$ and an upper limit of 0.32 on the hardness ratio. It is also associated to the radio source 112 in the VLA catalogue by \citet{Kellermann2008} with a flux density of $524\pm14\mu$ Jy and $390\pm11\mu$ Jy at 1.4 GHz and 4.8 GHz respectively, corresponding to a spectral index $\alpha=-0.2\pm0.1$. Although both these radio and X-ray analyses do not unambiguosly point to the presence of nuclear activity, \citet{Middelberg2011} detected it in the VLBI observations as a point-like source (S423) whose brightness temperature $2.2\times10^{6}~K$ clearly indicates an AGN. They estimate a 5GHz luminosity of $7.5\times10^{23}~W~Hz^{-1}$ typical of Seyfert galaxies. We argue that the emission lines in the BCG spectrum discussed by \citet{Afonso2006}, as the X-ray and radio emission, are also caused by the presence of an AGN because its SED is typical of an old galaxy with a passively evolving stellar population and negligible star-formation rate. Fitting a mixed stellar+AGN SED does not appreciably change the mass estimate, which indicates that the stellar population dominates the IR emission of the galaxy. This BCG is quite remarkable: it has a very high formation redshift ($z_{form}\sim5$), it is the most massive object in the GOODS South sample and among the most massive BCGs found at high redshift \cite[see, e.g.,][]{Stott2010}. Moreover, the BCG  is very bright ($K\simeq-26.7$ rest frame, referred to Vega) and it is $\sim 1.2$ mags brighter than the second brightest galaxy in the K band. According to \citet{Smith2010}, clusters hosting these dominant BCGs are quite rare even at lower redshift: they present a low degree of substructures and tend to be more centrally concentrated than average, suggesting that they formed at an early epoch.

We defined the fraction of red galaxies $f_{red}$ as the fraction of objects with $(V-Z)\geq 1.7$ at $M_{V}\leq-20.0$ considering both spectroscopic members of the cluster and photometric members as defined above. We considered galaxies at $r<0.6\times R_{200}$ to provide a comparison with the relation between velocity dispersion and fraction of E+S0 galaxies in high-z clusters by \citet{Poggianti2009}.  This is a conservative selection, because photometric members are all ``blue'' galaxies and part of them might be interlopers.  We obtain $f_{red}=74 ^{+2} _{-5}\%$, considering the uncertainty on the $R_{200}$ estimated from the full spectroscopic sample.  The fraction of red galaxies within the entire virial radius is $\sim 60\%$ (Fig.~\ref{fig:colmag}).  These values are considerably higher than the fraction of bright early-type galaxies ($\sim40\%$) found by \citet{Poggianti2009} in high-redshift ($z>0.5$) clusters with $\sigma_v\sim500-700~ Km/s$, but is consistent with what they find in nearby ones, suggesting that the galaxy population in \clgs~is significantly more evolved than in other distant clusters. 
Along with the high mass and luminosity of the BCG, the high fraction of red galaxies is an indication that \clgs~had formed at a very early epoch.

\section{X-ray properties}\label{xray}

\begin{figure}
\includegraphics[height=8cm]{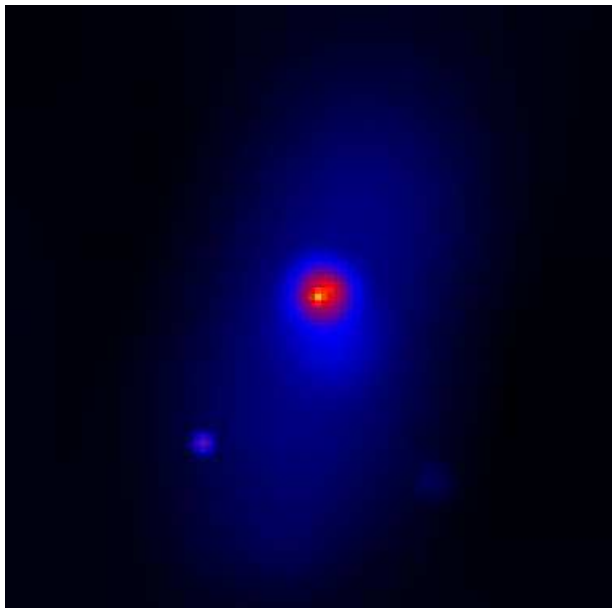}
\caption{Smoothed Chandra (0.5-1.5 keV) image of the extended emission of \clgs. The central point source is the BCG  \cite[ID=173 in the 2Ms catalogue by][]{Luo2008}. The side of the box is $\sim0.4~$Mpc.}
\label{fig:fig_clgs_X}
\end{figure}

In the combined 4Ms image we detect a faint extended X-ray source, centred on the position of the BCG  (Fig.~\ref{fig:fig_clgs_X}). After masking the BCG and three other X-ray point sources, we obtain $115\pm35$ counts (S/N=3.3) in the 0.5-1.5 keV band in an aperture of radius 27.6 arcsec ($\simeq0.2$Mpc) that apparently encloses all detectable flux. The S/N in the various bands is definitely too low to perform an accurate spectral fitting.  
However, the galaxy velocity dispersion in relaxed clusters is tightly correlated with the gas temperature because they both depend on the depth of the potential well. Following the $\sigma_v-T$ relation estimated by \citet{Xue2000}, we obtain a temperature $kT=3 \pm 1~$keV for our cluster, considering the velocity dispersion of all spectroscopic members. 
\begin{figure}
\includegraphics[height=8cm]{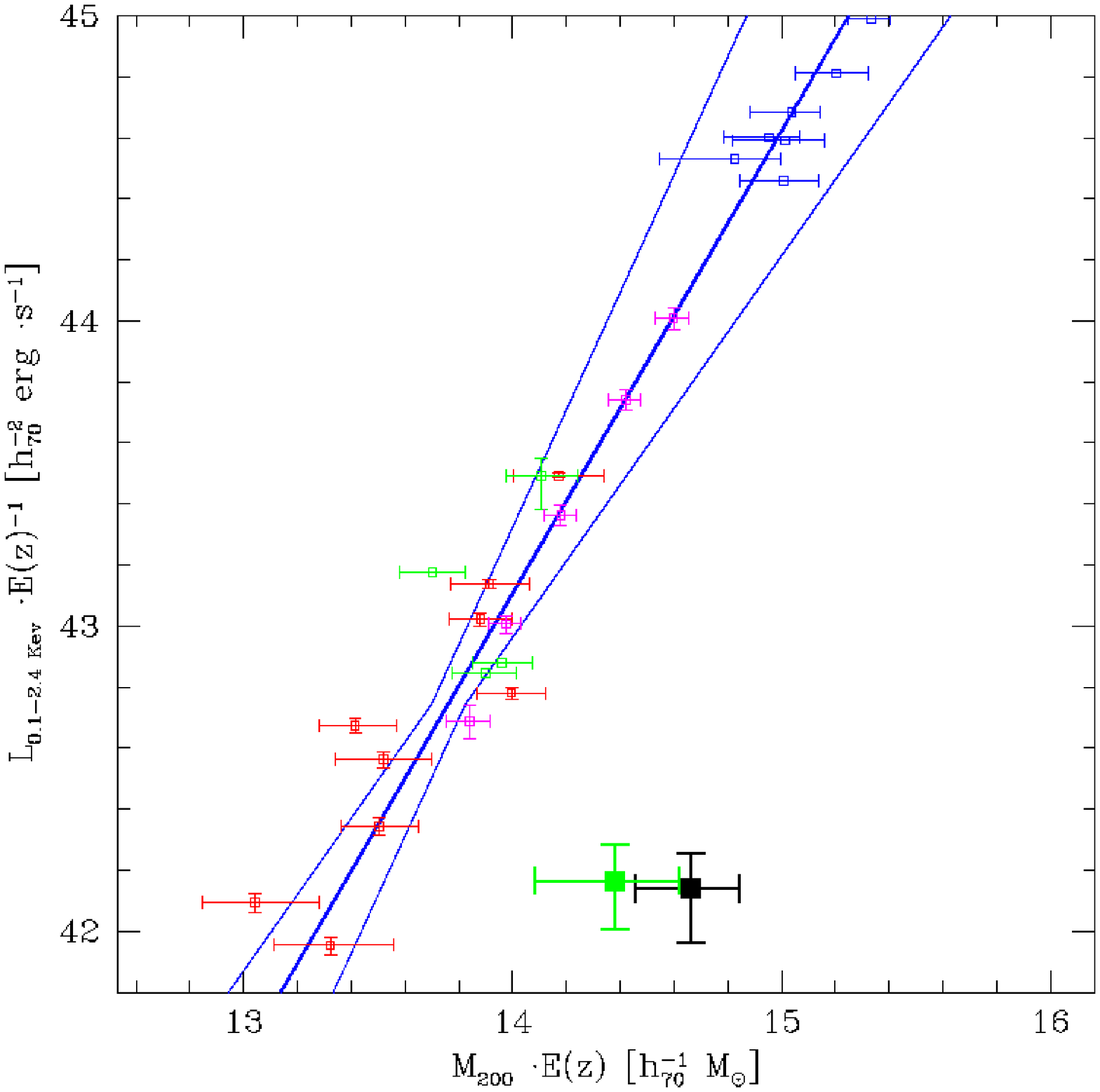}
\caption{Cluster \clgs~ in the $M_{200}-L_{0.1-2.4 keV}$ plane, compared with the scaling relation estimated by \citet{Leauthaud2010} (continuos lines) and with clusters from the literature: \citet{Leauthaud2010} (COSMOS, red empty squares), \citet{Berge2008} (CFHTLS, green), \citet{Hoekstra2007} (CFH12k, blue), \citet{Rykoff2008} (SDSS, magenta): the black filled square is derived from the full galaxy sample, the green one from red-sequence members only. $E(z)=H(z)/H_{0}$, all quantities were reported to the cosmology adopted in the present paper, \citet{Rykoff2008} masses were boosted by a factor 1.24 following \citet{Leauthaud2010}.}
\label{fig:fig_scaling}
\end{figure}
\begin{figure}
\includegraphics[height=8cm]{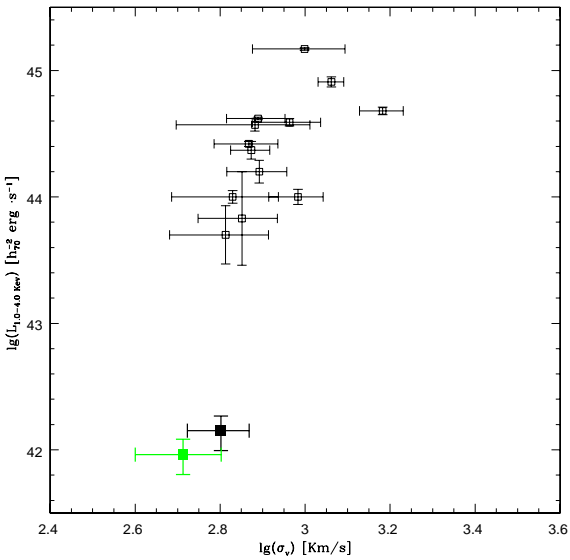}
\caption{Cluster \clgs~(black and green filled squares as in  Fig~\ref{fig:fig_scaling}) in the $\sigma_v-L_{1-4 keV}$ plane, compared with the high-redshift cluster sample of \citet{Andreon2008}.}
\label{fig:fig_sigmaL}
\end{figure}
The flux in the observed 0.5-2.0 keV band is $F_{0.5-2}=2.3 \pm 0.7~10^{-16}~erg~cm^{-2} ~s^{-1}$ when a thermal emission plasma model \cite[MEKAL,][]{Mewe1985,Mewe1986} with fixed metallicity and temperature ($Z=0.3 Z_{\odot}$, $kT=3 \pm 1~$keV) is applied. 
To convert the measured fluxes to rest frame total $L_X$ in the 0.1-2.4 keV band, we estimated K-corrections for the corresponding spectra. We corrected for the missing flux at larger radii following the procedure by \citet{Bohringer2004}: we integrated up to 12 core radii an isothermal beta-profile \citep{Cavaliere1976} with the parameters $\beta=0.4\cdot(kT/1 keV)^{1/3}=0.58^{+0.06}_{-0.07}$ and $r_c=0.07\cdot(kT/1 keV)^{0.63} \cdot R_{500}=0.1^{+0.04}_{-0.03}~Mpc$. 

We obtain $L_X=2.1 ^{+0.69}_{-0.64}~10^{42}~erg ~s^{-1}$. The velocity dispersion of red sequence galaxies implies a lower temperature range (1.4-3.0 keV) and a slightly shallower gas profile, yielding $L_X=2.2 \pm 0.68 ~10^{42}~erg ~s^{-1}$.

We compare the properties of \clgs~ with the latest estimate of the $M_{200}-L_X$ scaling relation by \citet{Leauthaud2010}. Their mass estimates come from a weak-lensing analysis of the X-ray sample detected by \citet{Finoguenov07} in the COSMOS Field, while X-ray total luminosities were computed assuming profile parameters as outlined above. 
The position of \clgs~ in the  $M_{200}-L_X$ plot is shown in Fig.~\ref{fig:fig_scaling}. Considering the virial mass estimated from the full spectroscopic sample  (black square and error bars), the $L_X$ is too low by a factor of $\gtrsim$20 with respect to the scaling relation measured from X-ray "bright`` galaxy clusters. The X-ray luminosity is discrepant by a factor of $\gtrsim10$ with the scaling relation, even considering the virial mass estimate $M_{200,red}$ obtained from the red galaxies only (green square in Fig.~\ref{fig:fig_scaling}).

We also show in Fig.~\ref{fig:fig_sigmaL} a comparison with the high-redshift cluster sample of \citet{Andreon2008} in the $\sigma_v-L_X$ plane. We assumed the same profile parameters as above to compute a total $L_X=1.42 \pm 0.43~10^{42}~erg ~s^{-1}$ in the rest frame 1.0-4.0 keV band, for $kT=3 \pm 1$keV. Once optical estimates obtained from red galaxies are considered (green square), the corresponding luminosity is lower than in the 0.1-2.4 keV band, since the $\sigma_{v,red}$ implies a lower temperature and thus a reduced emission in the 1-4 keV band. Again, cluster \clgs~ shows a much lower luminosity than high-redshift clusters with similar velocity dispersion.
 
\section{Discussion}\label{discussion}
\begin{figure*}
\includegraphics[height=10cm]{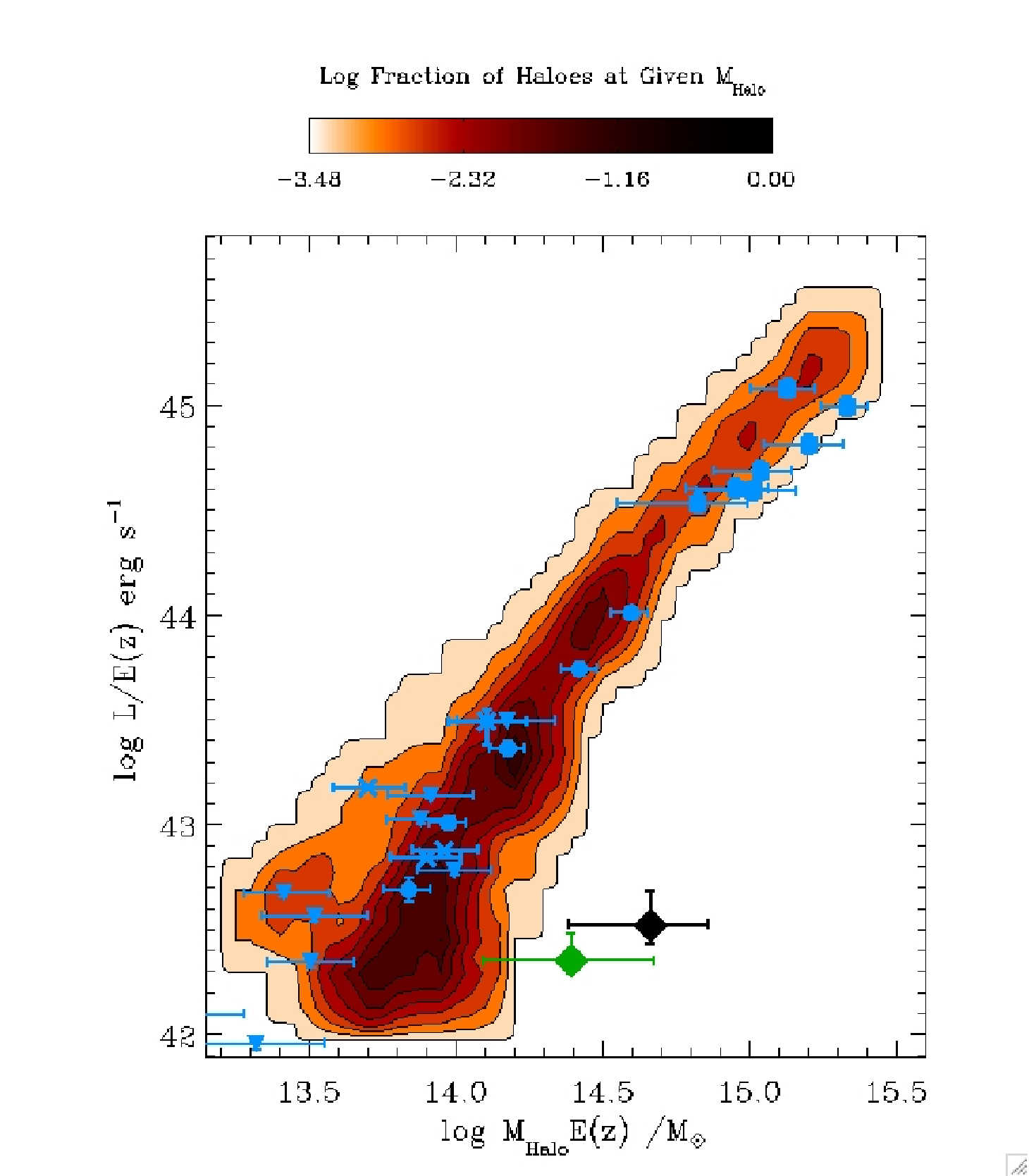}
\includegraphics[height=10cm]{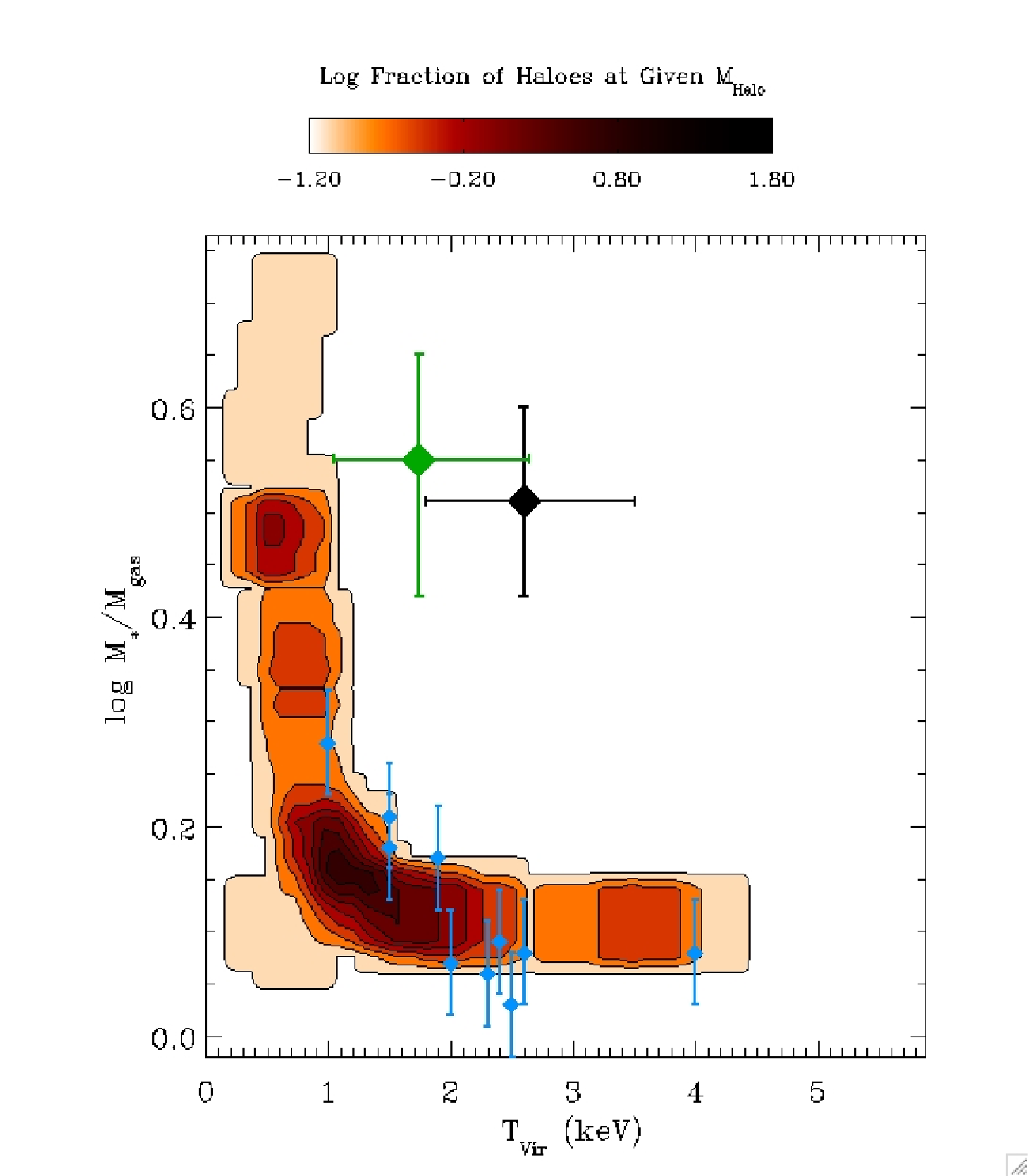}
\caption{Left: The fraction of clusters as a function of mass and X-ray luminosity in the semianalytic model (colour coded contours), along with the position of cluster \clgs~ in the $M-L_{X}$ plane: the black filled square is derived from the full galaxy sample, the green one from red-sequence members only (see main body for details). Blue symbols and error bars are the clusters from the literature represented in Fig.~\ref{fig:fig_scaling}. Right: The fraction of clusters as a function of virial temperature and $M_{star}/M_{gas}$ ratio in the semianalytic model (colour-coded contours), along with the relevant quantities for cluster \clgs~ (black and green squares) and for the clusters in the sample by \citet{Dai2010} (blue).}
\label{fig:fig_pera}
\end{figure*}
The huge discrepancy between the X-ray luminosity of \clgs~and of clusters with similar $M_{200}$ can be explained by a lower than expected gas mass, density, or temperature.

As shown by \citet{Cavaliere2002} and by \citet{Lapi2005}, the energy injection $\Delta E$ from SNe and AGNs can expand the cluster gas halo and expel a substantial fraction of its original mass. This implies a shallower profile (i.e. a lower $\beta$ parameter) and a reduced total X-ray emission.
The overall feedback effect depends on the merging history of the cluster, being related to the ratios $\Delta E/E$ between the injected energy and binding energy $E$ in each progenitor halo of the cluster. 
We exploited our semianalytic model \cite[described in detail in][]{Menci2006,Menci2008b} to estimate the scatter in the  $M_{200}-L_X$ and to assess whether a strong feedback is capable of lowering the X-ray luminosity at the level observed in \clgs. This model successfully reproduces the observed properties of both galaxies and AGNs across a wide redshift range \cite[e.g.][]{fontana06,Menci2008,Calura2009,Lamastra2010}, and it is the same one adopted in \citet{Cavaliere2002} and in \citet{Lapi2005} to provide a comparison with the properties of X-ray detected clusters. 

\begin{figure*}
\includegraphics[height=10cm]{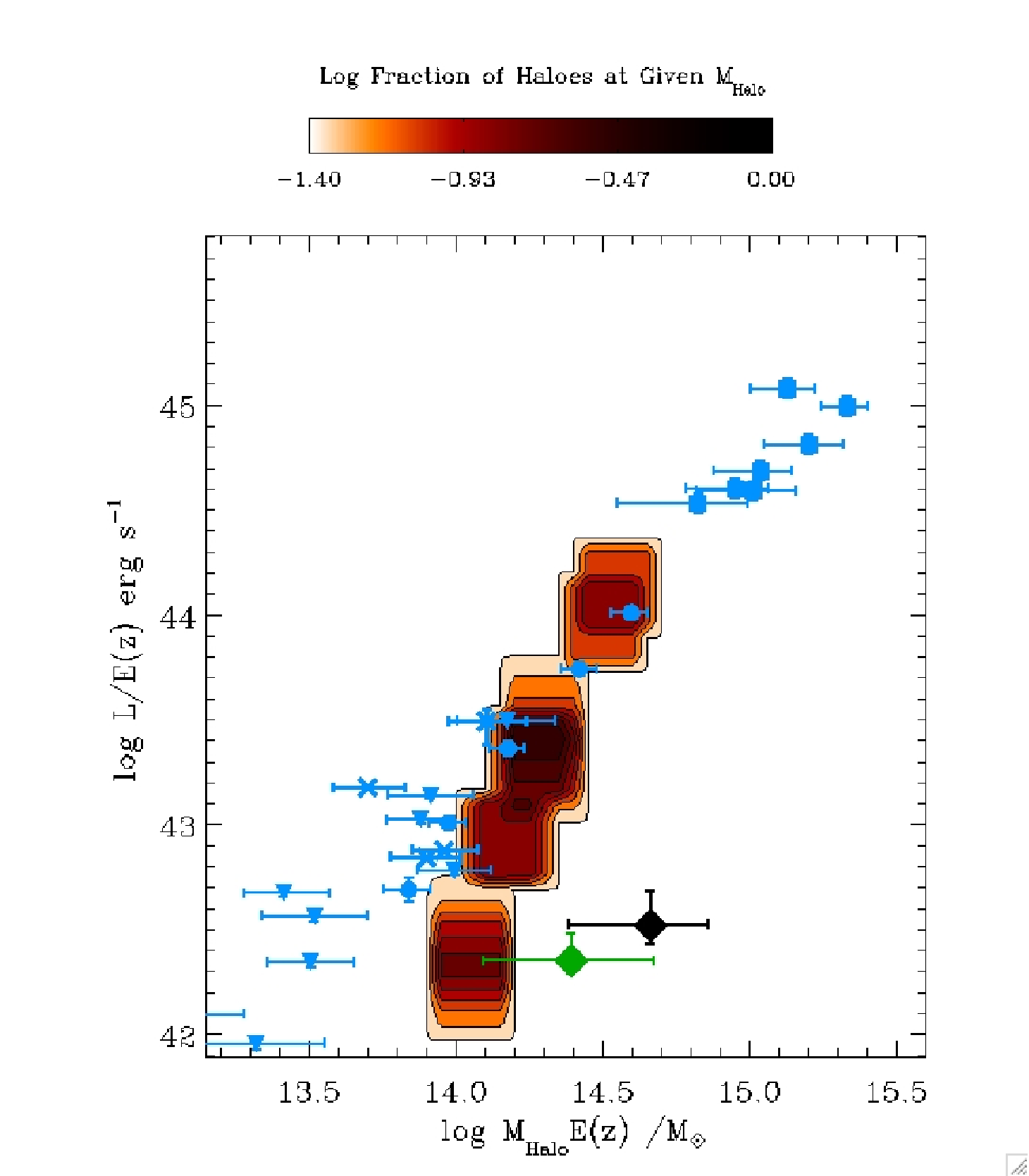}
\includegraphics[height=10cm]{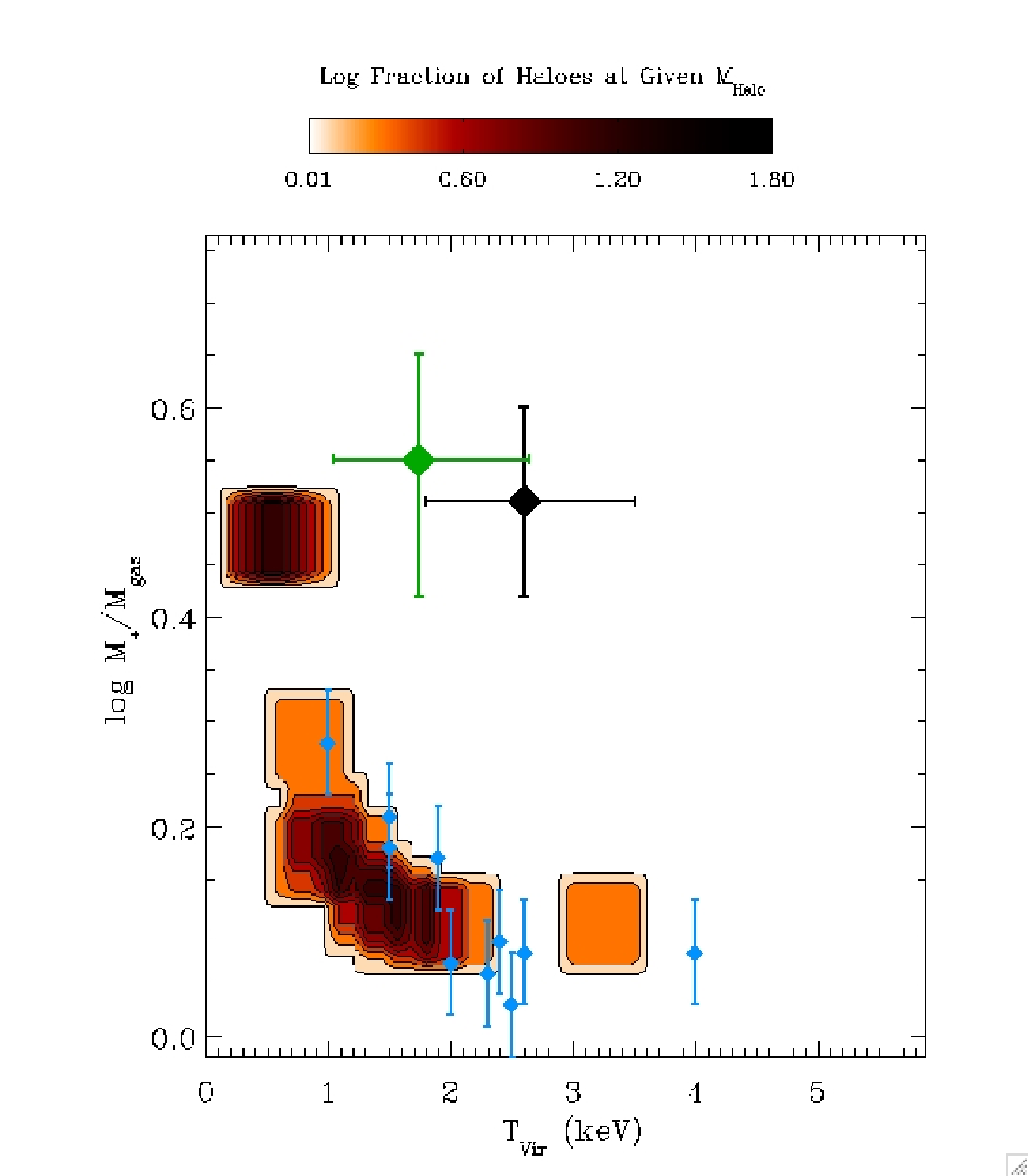}
\caption{Same as Fig.\ref{fig:fig_pera} but for clusters with $f_{red}>60\%$ and $M_{star}>5\times10^{12}~M_{\odot}$.}
\label{fig:fig_pera_allcut}
\end{figure*}
We consider $10^5$ simulated halos with M$\gtrsim2\times10^{13}~M_{\odot}$: in Fig.~\ref{fig:fig_pera} we show their distribution in the $M-L_X$ plane as colour-coded contours. The model agrees with the observed position of ''X-ray bright`` clusters in the diagram (blue points), but it shows a large scatter in the $M-L_X$ relation at low masses owing to the effect of feedback. To provide a more meaningful comparison with the simulated clusters, the total X-ray luminosity of \clgs~was re-determined assuming a broader profile with $\beta=0.45$, which is the value corresponding to the strongest non-thermal energy injection in the model, and integrating up to $R_{200}$. This shallow profile implies a luminosity that is slightly higher than in Fig.~\ref{fig:fig_scaling}, although still highly discrepant with the range of luminosities predicted for structures of this virial mass (black point and error bars).

In Fig.~\ref{fig:fig_pera} we also show the $M_{star}/M_{gas}$ ratio as a function of the virial temperature for \clgs~(black) and for the clusters reported in \citet{Dai2010} (blue points), along with the relevant distribution in the simulated clusters. The total gas mass within $R_{200}$, $M_{gas}= 6.5 ^{+1.1}_{-1.3} ~ 10^{12}~M_{\odot}$, was computed assuming $\beta=0.45$ as before, and pure bremsstrahlung emissivity; we obtain $M_{star}/M_{gas}=0.55^{+0.1}_{-0.13}$. An even higher ratio ($M_{star}/M_{gas}\sim0.8$) is found assuming average profile parameters as in Sect.~\ref{xray}. Notably, the model reproduces the $M_{star}/M_{gas}$ ratios measured by \citet{Dai2010} quite well, but ratios as high as for \clgs~are not found in simulated clusters of similar virial temperature.

When considering the estimates based on red galaxies only (green points and error bars in Fig.~\ref{fig:fig_pera}), \clgs~appears to be marginally compatible with the lower luminosity simulated clusters in the $M-L_X$ plane, and its $M_{star}/M_{gas}$ ratio is barely compatible with those found in the cooler simulated structures.

Nonetheless, the analysis of the connection between the model predictions for the ICM emission and for the properties of cluster galaxies allows us to derive interesting clues on the possible physical origin for the peculiarity of \clgs. 
As noted in Sect.~\ref{stellarmass}, \clgs~shows a high fraction of red, bright galaxies, which is comparable to the ratio observed only in very massive clusters at $z>0.5$. 

We thus considered only the subsample of simulated clusters (8 \% of the total) with a red fraction as high as \clgs~ ($f_{red}>60\%$) and a high stellar mass ($M_{star}>5\times10^{12}~M_{\odot}$). We show in Fig.~\ref{fig:fig_pera_allcut} the same plots as in Fig.~\ref{fig:fig_pera} for these highly evolved clusters. These objects show a scatter of a factor of 10-20 in X-ray luminosity at fixed total mass. A higher fraction of objects with a high $M_{star}/M_{gas}$ ratio is found among them. This is due to their formation history: their many red massive galaxies formed in biased environments at high redshift, and they likely host supermassive black holes whose feedback impact is particularly effective in expelling the ICM from the potential well. While these highly evolved clusters appears to be closer to \clgs~than ``average'' high-redshift ones, the remaining discrepancies between them and \clgs~indicates  that the model might still underestimate the effect of feedback. In particular, $M_{star}/M_{gas}$ ratios as high as in \clgs~ are not found among simulated clusters of similar $T_{vir}$, suggesting that the model underpredicts the fraction of gas mass expelled because of the effect of feedback. 

This scenario can be empirically tested by the analysis of a larger sample of high-redshift clusters through looking for a correlation between anomalous X-ray emission and the presence of a high fraction of evolved galaxies within the virialized region. However, a deeper understanding of the connection between the effects of AGN feedback on galaxies and on the ICM is required to provide an exahustive answer on this point.  In this respect, the observed properties of \clgs~ discussed in Sect.~\ref{stellarmass} are consistent with an early formation epoch and thus with a feedback impact in its past history greater than in other clusters of the same mass, although the feedback should have been even more effective than what is predicted by the model. We note that our model only includes the mechanical feedback from the blast wave triggered during the bright QSO phase of AGN activity. As shown by \citet{Giodini2010}, radio-active galactic nuclei can also unbind a significant fraction of the ICM in galaxy groups. It is thus possible that including this additional feedback along the evolutionary history of simulated clusters could reconcile the observation of X-ray underluminous structures such as \clgs~with the model predictions.

\subsection{Deviations from virial equilibrium and possible failures of optical estimates}
Albeit a larger than expected impact of feedback provides a suggestive and physically grounded scenario for the existence of underluminous clusters, other possible explanations have been proposed. 
A low X-ray luminosity can also be explained if the gas is not virialized and far from the high temperature needed to emit the expected $L_X$. 
Indeed, it has been argued  that X-ray underluminous clusters might still be in the process of formation \cite[e.g.][]{Bower1997,Rasmussen2006,Popesso2007}. While this might be the case for some clusters where the velocity distribution strongly deviates from a Gaussian \cite[as in][]{Rasmussen2006}, we argue that this is no likely explanation for \clgs. The analysis of the region within $R_{200}$ discussed in Sect.~\ref{virstatus}, and the galaxy properties discussed in Sect.~\ref{stellarmass}, provide strong evidence that the cluster has already relaxed. An unvirialized status of the gas within the inner region would be in contrast with the simple theoretical expectation that gas and galaxies need comparable timescales to reach virial equilibrium \cite[e.g.][]{Gunn1972}. On the other hand, the unrelaxed nature of the external regions found in X-ray underluminous Abell clusters by \citet{Popesso2007} might not be directly related to the status of the cluster core, because the time elapsed from the first collapse of the structure is usually shorter than the timescale needed by distant galaxies to relax in the cluster potential well \cite[e.g.][]{sarazin,voit2005}. The detection of collapsing systems and of clusters with substructures or complex morphologies, but which display a high X-ray emission \cite[e.g.][]{Valtchanov2004,Bourdin2010}, further indicates that a low X-ray luminosity is more probably related to the evolutionary history of the ICM than to the virialization status.

As noted by \citet{Bower1997}, clusters classified as ''X-ray underluminous`` might be poor systems where optical mass estimates fail because they are embedded in a filament observed along the line of sight. In this case, we expect blue, star-forming galaxies to be mostly part of the environment surrounding the cluster. We exclude this hypothesis in our case, because the population of blue galaxies in \clgs~ show a velocity dispersion that is consistent with being Gaussian (Sect.~\ref{virstatus}), as expected if they have undergone relaxation in the cluster potential well. Although mass estimates might suffer of systematic uncertainties even when cluster members are properly selected (as extensively discussed by B06), \clgs~is still discrepant with known scaling relations even considering the mass estimated from red-sequence members only, which tend to underestimate rather than overstimate the true dynamical mass of the cluster.

These considerations lead us to conclude that the most likely explanation for the existence of X-ray underluminous clusters lies in a large scatter in the $M-L_X$ relation owing to feedback effects, as suggested by the properties of simulated clusters. The remaining discrepancy between model predictions and the properties of \clgs indicates that feedback effects might be even stronger than what is contemplated by our model. However, we underline that only a multiwavelength analysis of larger cluster samples will provide a clear answer to this question.

\section{Summary and conclusions}\label{summary}
We have presented a spectroscopic and X-ray analysis of the cluster \clgs~at z=0.7347 in the GOODS-South field, based on the newly released Chandra 4MS observations and on the large amount of public spectroscopic data available.  We estimate a velocity dispersion $\sigma_v=634 \pm 105~Km/s$ from 44 spectroscopic members within 1 Mpc from the BCG, and $\sigma_{v,red}=516\pm118~Km/s$ considering only 20 red sequence members.  From these velocity dispersions, following standard procedures, we estimate  $M_{200}=3.07 ^{+1.57} _{-1.16}~10^{14}~ M_{\odot}$, ($R_{200}=1.06 ^{+0.14} _{-0.15}~Mpc$), and $M_{200,red}= 1.6^{+1.16} _{-0.8} ~\cdot 10^{14} ~M_{\odot}$, ($R_{200,red}= 0.86 ^{+0.16} _{-0.17}~Mpc$), respectively. Galaxies within the virial radius have a velocity distribution that is consistent with a Gaussian according both to classical (K-S test, skewness, kurtosis) and to robust one dimensional tests (asymmetry and tail indices). There is no evidence for substructures within the virial core according to the Dressler-Schechtman test ($P(\Delta)>0.1$). These estimates are consistent with virial equilibrium within $\sim$1 Mpc from the cluster center. We measure a total stellar mass of $M_{star}=3.3 \pm 0.5~10^{12}~ M_{\odot}$ from both spectroscopic and photometric members of the cluster. The BCG is among the most massive found in high-redshift clusters, contributing about one third of this value; it is also a radio and X-ray detected AGN. Cluster \clgs~has a high fraction of red galaxies $f_{red}\simeq74\%$ at $M_{V}\leq-20.0$, which is unusually high when compared with the fraction of early-type galaxies of these luminosities in similar high-redshift clusters \citep{Poggianti2009}. The high fraction of red galaxies and the properties of the BCG suggest that the cluster is in an evolved status and was formed at an early epoch.

We detected an extended X-ray source centred on the BCG, whose total X-ray luminosity in the range 0.1-2.4 keV ($L_X\simeq2~10^{42}~erg ~s^{-1}$) is different by a factor of $\sim10-20$  with the latest estimates of the observed $M-L_X$ relation considering either the $M_{200}$ and $M_{200,red}$ values for the virial mass. Cluster \clgs~ thus appears to be one of the so-called ''X-ray underluminous`` clusters reported in the literature.

We explored possible explanations for this discrepancy.  We argue that it is unlikely that the ICM has not yet virialized, because the virial core appears to be relaxed and free of substructures, and gas is expected to heat to the virial temperature on a short timescale.
We find it more likely that the cluster gas has been heated and in part was expelled from the potential well because of strong AGN feedback. This is suggested by the comparison with our semianalytic model, which shows that clusters with an evolved galaxy population like \clgs~present a large scatter in X-ray luminosity at fixed mass. However, the X-ray emission of \clgs~is still discrepant with model predictions when considering the $M_{200}$ estimated from the full galaxy sample and it is only marginally compatible with simulated clusters when considering the $M_{200,red}$ estimate. The remaining difference with respect to simulated clusters might indicate that the model still underestimates the effect of feedback on the ICM.

The apparent relaxed status and the properties of \clgs~indicate that ''X-ray underluminous`` clusters of this kind are real structures and that the population of galaxy clusters is more heterogeneous than what appears in X-ray flux limited samples.
We note that this scenario qualitatively agrees with the reported discrepancies between optically and X-ray selected cluster samples. Indeed, we expect optical detections to effectively recover the fraction of clusters where feedback effects have been stronger, thus having a lower luminosity for their temperature \citep{Bignamini2008}, shallower profiles \citep{Rykoff2008} and higher stellar-to-gas mass ratios \citep{Dai2010} with respect to X-ray selected ones. The recently published early results of Planck SZ observations of galaxy clusters also agree with this scenario. \citet{Planck02} presented a sample of massive SZ-detected clusters at intermediate redshift that are underluminous for their mass and have flatter density profiles with respect to X-ray selected ones. Moreover, \citet{Planck01} found that the relation between the SZ signal $Y_{500}$ (proportional to gas mass and temperature) and cluster richness $N_{200}$ for objects in the MaxBCG cluster catalogue \citep{Koester2007} has a lower normalization and larger scatter than expected, probably because of the presence of a population of X-ray underluminous clusters seen only in the optically selected catalogue.
This scenario can only be confirmed by looking for evidence of evolved galaxy populations and signs of early formation epochs in distant, optically selected clusters that are discrepant with scaling relations measured in X-ray selected samples. If this hypothesis is correct, it will be necessary to adopt multiwalength strategies in detecting and analysing high-redshift clusters, both to study their formation and evolution and to constrain cosmological parameters from their mass function.   

 \begin{acknowledgements}
MC and LP  acknowledge financial support from ASI contract I/016/07/0.
 \end{acknowledgements}
\bibliographystyle{aa}

\end{document}